\documentclass[proof]{WileyASNA-v1}

\articletype{Original Article}%

\received{31 Aug 2018}
\revised{8 Oct 2018}
\accepted{10 Oct 2018}

\begin{document}

\title{The Vast Potential of Exoplanet Satellites for High Energy Astrophysics}

\author[1,2]{Krista Lynne Smith*}



\authormark{K.L. Smith}

\address[1]{\orgdiv{KIPAC at SLAC}, \orgname{Stanford University}, \orgaddress{\state{California}, \country{United States of America}}}

\address[2]{Einstein Fellow}


\corres{* \email{klynne@stanford.edu}}

\presentaddress{2575 Sand Hill Rd., Menlo Park, CA 94025}

\abstract{The photometric precision, monitoring baselines and rapid, even sampling rates required by modern satellites designed for detecting the signal of transiting exoplanets are ideally suited to a large number of applications in high energy astrophysics. I will exemplify this by discussing results for AGN from \textit{Kepler} and summarizing other high energy results from Kepler/K2. These precision instruments are currently underutilized for high energy applications despite their great potential, due in part to complex systematics affecting the data. I will summarize these effects and mitigation approaches, and conclude by discussing how the recently-launched Transiting Exoplanet Survey Satellite mission will differ from \textit{Kepler}/K2 in ways significant to the high energy community.}



\maketitle


\section{Introduction}\label{sec1}

In 2009, NASA launched the \textit{Kepler} spacecraft with a highly focused goal: stare unblinking at a patch of the sky near the Milky Way to detect the transiting signal of planets around other stars. In many ways the successor of the European CoRoT mission, \textit{Kepler}'s goal required a revolutionary new take on optical time domain studies. Where ground-based timing had for decades been affected by diurnal and seasonal gaps, irregular sampling and often poor photometric precision, \textit{Kepler}'s objectives would not tolerate any of these conditions. The result was an unprecedented spacecraft designed to accumulate regularly-sampled, extremely high precision light curves with 30-minute optical cadence over a baseline of many years. NASA solicited guest observer proposals for the use of this incredible instrument for non-exoplanet applications so long as targets remained within the fixed 115~deg$^{2}$~field-of-view (FOV). The mission was immensely successful at its planned goal, but \textit{Kepler}'s long baseline, rapid cadence, even sampling, and impeccable photometry also opened a new window in the optical time domain with potential applications for a huge variety of astrophysical targets. Perhaps the most prolific use has been for asteroseismology, igniting a new golden age in stellar structure and evolution. Far slower and less prolific were studies related to high-energy astrophysical phenomena, especially extragalactic applications. 

For decades, X-ray satellites had been providing high-cadence, evenly sampled light curves of active galactic nuclei (AGN), X-ray binaries (XRBs), cataclysmic variables (CVs), and other accreting sources. The X-ray light curves led to a wave of insights into the accretion of matter on to black holes, neutron stars, and white dwarfs. In the case of AGN, rapid X-ray timing has been invaluable, providing clues to the geometry and nature of accretion flows very near to the supermassive black hole itself \citep[e.g., reverberation studies, ][]{Uttley2014}. Furthermore, comparing simultaneous variability in optical and X-rays may offer tantalizing insights, but optical ground-based timing is often not directly comparable to the space-based X-ray light curves due to gaps, irregularity, and slow cadence. With \textit{Kepler}, optical light curves were finally on similar footing to X-ray light curves, and the same physical processes could be potentially studied. Because accretion disks are too small to be directly imaged with any current or upcoming facilities, variability provides a rare direct observational probe. In AGN, the vast majority of the optical radiation comes from the accretion disk \citep{Shields1978}, so the behavior of this light offers a valuable glimpse into the disk's physical processes. However, due partly to a dearth of intersectional marketing of an exoplanet satellite to the high energy community, the prime \textit{Kepler} mission was seriously underutilized for these applications. Additionally, since the spacecraft's primary goal was intensely focused, the data delivered in the initial releases were poorly suited to many high energy applications, slowing progress and requiring the development of custom software by investigators.

In this proceeding, I will discuss progress and results on AGN, provide an overview of the challenges posed by the data and solutions from the literature, and briefly review other non-AGN high energy results that have come out of the \textit{Kepler} mission and its reincarnation, K2. I will conclude by discussing the potential for such applications using the recently-launched Transiting Exoplanet Survey Satellite (TESS), with additional remarks on and the future European CHaracterising ExOPlanet Satellite (CHEOPS) and PLAnetary Transits and Oscillations of stars (PLATO) missions.

\section{Kepler: AGN Applications}\label{sec2}

The potential physical insights into AGN physics from space-based optical light curves depend on the AGN type; however, the primary (though not exhaustive) observables are: 1) the frequency and amplitude of flares, dips or events in the light curve itself, 2) the high-frequency slope of the power spectral density function (or discrete Fourier transform) of the light curve and whether this corresponds to predictions from theoretical models, and 3) features in the power spectral density function that deviate from a simple single power law, such as breaks that would indicate a characteristic timescale or features that would indicate periodicity. 

The first major hurdle to accomplishing AGN science with \textit{Kepler} was the scarcity of AGN in the pre-defined FOV. The catalogued AGN known in the field included only the Seyfert 1 II~Zw~229-015 and the radio galaxy Cygnus A. Various methods were undertaken to find new AGN, including archival infrared photometric colors \citep{Edelson2012}, archival compact bright radio cores \citep{Wehrle2013}, and a new X-ray survey \citep{Smith2015}.  A few dozen AGN were discovered via these different methods, including blazars and Type I and II Seyferts. 

The first AGN results using \textit{Kepler} light curves were published by \citet{Mushotzky2011}, who measured power spectral density slopes in four Type I Seyfert AGN with one quarter ($\sim$3 months) of data and found that they were steeper than slopes in ground-based timing studies (i.e., there was more variability power at high frequencies). Steep slopes are inconsistent with predictions from the analytic damped random walk \citep[DRW;][]{Kelly2009} model for accretion disks but more in line with early predictions from MHD turbulence modeling of accretion rate variations. The inconsistency of \textit{Kepler} AGN light curves to the damped random walk was also found in a larger study by \citet{Kasliwal2015a}. The increased photometric precision and fine sampling likely contributed to the tension with ground-based studies. 

The following year, \citet{Carini2012} published a paper focused on II~Zw~229-015, reporting the discovery of a characteristic variability timescale of order 44 days or 90 days, depending on the model used, and similarly steep slopes to \citet{Mushotzky2011}. One year after that, two papers were written on blazar-type AGN: \citet{Edelson2013} published a detailed analysis of the BL Lac W2R~1926+42, and \citet{Wehrle2013} published work on three flat-spectrum radio quasars (FSRQs) and one radio-loud Type 1.5 AGN. The \citet{Edelson2013} light curve was ground-breaking in duty cycle and precision for blazars, and results included a power spectrum poorly fit by broken or single power law models, a possible non-linear relationship between flux and rms variability, and numerous rapid flares at more than 25\% of the base intensity. \citet{Wehrle2013} did not find such intense flares in their FSRQ sample, and reported relatively shallow power spectral slopes comparable to ground-based studies; this result differed from the \citet{Mushotzky2011} result of steeper slopes, albeit for traditional Seyfert-type AGN. In 2014, there were two follow-up studies of the AGN already published: \citet{Edelson2014} performed their own analysis of II~Zw~229-015 with a different data reduction process and found a characteristic variability timescale of $\sim$5 days rather than 44, and \citet{Revalski2014} re-analyzed the same four objects as \citet{Wehrle2013} but with three added quarters ($\sim9$~months) of monitoring, confirming the shallow power spectral slopes but detecting new significant flaring behavior in one object. \citet{Aranzana2018} analyzed a very large sample of \textit{Kepler} and K2 AGN and found a wide range of power spectral slopes. Finally, \citet{Smith2018} performed a detailed analysis of 21 AGN in the original \textit{Kepler} field, finding again a wide range of power spectral slopes but also reporting six objects with characteristic variability timescales.

All of these results were important, leveraging the \textit{Kepler} data to explore a parameter space previously inaccessible to optical timing and producing a complementary set of results to ground-based efforts. However, the data were affected by complex and numerous systematic effects driven by spacecraft behavior. Unfortunately, many of these effects are especially adversarial to science applications in accretion, since accretion-driven variability is typically a stochastic red-noise process with long- and short-term behavior that is easily mimicked by the artificial spacecraft trends. Most studies were aware of these effects, and each took a different approach to mitigation, becoming naturally more sophisticated over the years. The next section details the major systematics and behaviors that deleteriously affect many high energy sources.

\subsection{Kepler and K2: Data and Products}\label{dataproducts}

The original \textit{Kepler} mission pointed at a 115 deg$^2$~field of view near the constellation Cygnus from $2009-2013$ \citep{Borucki2010}. It has one single, broad, white-light bandpass and monitored a large sample of candidate exoplanet host stars as well as guest-observer targets. Targets could be monitored at either long cadence (30 minutes) or short cadence (1 minute). \textit{Kepler} data can be downloaded as calibrated light curves or as target pixel files (TPFs). It is not wise to use the calibrated light curves for AGN or other applications where stochastic or red-noise variability is expected (see the next Section). Our work suggests that starting with the TPFs is essential for understanding what effects are at work in the data and for approaching their mitigation. The TPFs allow you to define your own extraction aperture, and include both corrected (Pre-search Data Conditioning or PDC) and uncorrected (Simple Aperture Photometry or SAP) data columns for each pixel. The final type of \textit{Kepler} data product is the full-frame images (FFIs): snapshots of the entire FOV taken every 30 days during the full course of the prime mission (2009-2013). While objects with 30 or 1 minute monitoring required proposals, one can construct a 30-day cadence light curve out of the FFIs for any sufficiently bright target in the entire FOV.

After the failure of \textit{Kepler's} redundant reaction wheel in 2013, the mission was redesigned as K2, which has the same monitoring cadences but a different FOV situation. Instead of staring at a fixed point for years, the spacecraft can monitor a pre-defined patch of the sky on the ecliptic plane for approximately 70 days before moving on to another patch. For K2 data, user-created light curves are also an option. There are numerous community-built pipelines available for both K2 and \textit{Kepler} data. For example, the \textit{Everest} pipeline uses pixel-level decorrelation for de-trending \citep{Luger2016} and the \textit{K2P}$^{2}$~pipeline algorithmically defines custom extraction apertures even in crowded fields \citep{Lund2015}. A list of community-delivered pipelines is available at \url{http://lightkurve.keplerscience.org/other_software.html}.

\subsection{Caveats from Kepler Data}\label{caveat}

It is extremely important to not use the pre-corrected light curves for non-exoplanetary science applications. In addition to the discussions in most of the papers referenced in this work, \citet{Kasliwal2015b} analyzed the effects of the spacecraft systematics in detail for II~Zw~229-015 specifically and concluded that the \textit{Kepler} AGN light curves require re-processing before any comparison with theoretical models can be done. The most important factors affecting the \textit{Kepler} light curves can be divided into two main categories: those that are easy to recognize or are flagged in the data and are due to well-understood spacecraft requirements, and those that are more insidious and may or may not be flagged. Provided here is a list of each type of effect, and how the problem has been approached. 

\subsubsection{Effects Due to Well-Understood Spacecraft Requirements}
\begin{itemize}
\item Quarterly discontinuities: In order to maintain the solar panels' orientation, \textit{Kepler} was rotated 90 degrees approximately every 3 months. This rotation placed different CCDs on each position within the field of view each quarter. Because each CCD is subject do variations in quantum efficiency, etc, there are large but predictable flux discontinuities in the data at quarterly intervals. Some authors have chosen to ignore this problem by studying only single quarters \citep[e.g.,][]{Mushotzky2011}, while some have attempted to stitch the quarters together either multiplicatively or additively. Ideally, even patchy simultaneous monitoring from the ground enables the quarters to be confidently stitched using the ground-based data as a guide to the true offsets \citep[e.g.,][]{Carini2012, Edelson2014}. However, this has only been possible for II~Zw~229-015. In K2 data, these discontinuities do not exist because no rolls are performed during each monitoring campaign.
\item Earth-point gaps: Each month, the spacecraft must point at Earth to transmit data. Cadences affected are flagged and gaps typically last $\sim$1 day. One can choose to interpolate over the gaps either linearly or using more sophisticated techniques like the LOWESS method - \citet{Smith2018} found that both methods resulted in power spectra with practically identical slopes and properties. \citet{Wehrle2013} filled gaps by fitting a third-order polynomial to data on either side and adding noise characterized by the local standard deviation. Alternatively, one can leave the gaps in the data and use different methods than Fourier transforms to obtain power spectra or their equivalents, such as structure functions or Lomb-Scargle periodograms, as done in \citet{Aranzana2018}.
\item Gaps from other flagged sources: Finally, there are numerous cadences with unique quality flags due to cosmic ray detections, attitude tweaks that result in false variability, etc. Usually these gaps are only one or two cadences in length ($\sim$1 hour) and can be safely interpolated over.
\end{itemize}

\subsubsection{Subtle Effects}
\begin{itemize}
\item Rolling Band Noise: Also known as Moir\'{e} pattern noise, this is an additive effect due to amplifier oscillations in the electronics that ripples across the detector, causing typically high-frequency variations that can often resemble the stochastic behavior of accreting sources. In \textit{Kepler} data releases DR25 and later, there is a column (``RB\_LEVEL") that identifies whether or not the rolling band artifact is present in a given cadence. Currently, work is underway at NASA in collaboration with AGN scientists to potentially mitigate the rolling band artifact, but right now, it is safest to remove the affected cadences or discard objects with large ranges of cadences affected by the rolling band (or, treat them with gaps using, e.g., the structure function).
\item Differential Velocity Aberration (DVA): As the spacecraft orbits the sun at L2, relativistic effects result in a subtle drift that causes a target to move very slowly across the CCD. Since \textit{Kepler} performs its light curve construction using aperture photometry, this can cause the source to drift until some of its flux departs the extraction aperture, causing an artificial drop in the light curve (or an artificial rise, in the reverse case). This is especially relevant for AGN (and perhaps tidal disruption events and supernovae) - phenomena that occur in extended host galaxies. The \textit{Kepler} default apertures were based on the assumption that the target is a point-source, and so are often too small to encompass the full extended source and exclude pixels with light from the host. When the drift occurs, more or less of the host galaxy starlight is included in the aperture, causing artificial rising and falling of the light curve. One approach to this problem is to use the target pixel files (TPFs) to create a very large extraction aperture. This way, the source can drift around and not depart the aperture. There are two main drawbacks to this approach: 1) if the source is in a crowded field, it is likely that any suitably large aperture to capture the full drift will also begin to approach other nearby sources that may then drift into it, and 2) much more background noise is included in the light curve. \citet{Aranzana2018} took the approach of modeling the DVA and determined that de-trended light curves were quite similar to real light curves. \citet{Smith2018} found that de-trending did not introduce spurious characteristic timescales, but the degree of the effect on power spectral slope varied dramatically. Whether and to what extent to de-trend will depend sensitively on the user's science application.
\item Thermal Focus Changes: Thermal fluctuations in the photometer cause the focus to change slowly, often mimicking the long-term, low-frequency trends caused by DVA. In addition, following an Earth-point event there is a strong thermal gradient that affects the focus and often settles after 2-3 days. These periods are not flagged in the data; in \citet{Smith2018} we chose to discard 150 cadences after each Earth-point. Alternatively, one can attempt to model the effect and remove it, but this requires assumptions about the underlying variability that often cannot be verified. It is important to note that DVA and focus changes are often superimposed upon one another, and can be challenging to disentangle. One approach is to use the co-trending basis vectors (CBVs) provided by the \textit{Kepler} mission to de-trend the data. The CBVs are a set of sixteen orthonormal functions that represent correlated features from a reference ensemble of stellar light curves, based on the idea that any correlated variability across many stars must be spacecraft-induced. However, applying these vectors to stochastically-varying light curves can seriously over-fit the data and remove true variability. An extensive discussion of the degree to which this occurs can be found in the appendices of \citet{Smith2018}; in summary, we chose to apply only the first two CBVs. This choice will vary based on the desired science application.  Current work is also underway towards modelling the long-term trends in K2 data as differential focus errors that cause bright and faint pixels to behave differently \citep{OBrien2018}.
\end{itemize}

\section{Stellar High-Energy Sources}

\subsection{Cataclysmic Variables}

Cataclysmic variables are binaries in which matter from a main sequence star is accreting onto a white dwarf. There are ten known CVs in the original \textit{Kepler} FOV, with two bright, well-known objects: V1504 Cyg and V344 Lyr. Both of these are of the dwarf nova sub-class of CV known as SU UMa type stars. For decades, these stars have been known to exhibit both normal outbursts and super-outbursting phases, with periods between super-outbursts referred to as supercycles. Testing theoretical models for the cause of superoutbursts and the duration of their supercycles requires well-sampled data of even the quiescent period, with good characterization of the normal outbursts. Until \textit{Kepler}, normal outbursts were rarely monitored for most SU UMa stars and ground-based all-sky surveys had cadences of approximately daily, too coarse to study orbital variations of CVs (typically a few hours). \textit{Kepler} provided unprecedented sampling of the entire dwarf nova cycle, including detailed profiles of the superoutbursts and clear morphologies of the ``superhumps" (periodic variability of the superoutburst phase, with a period a few percent longer than the orbital period).  \citet{Still2010} discovered superhumps in V344 Lyr that took place during quiescence rather than outburst - previously an elusive phenomenon. This discovery poses a significant challenge to the thermal-tidal instability model for mass transfer. \citet{Cannizzo2012} studied V344 Lyr and V1504 Cyg with longer light curves, and further rule out the thermal-tidal model by finding that normal outbursts increase to a local maximum and then decrease before the next superoutburst, where the model predicts that normal outbursts will get gradually stronger until the next superoutburst occurs. \citet{Dai2016} studied 15 CVs monitored in K2's first two campaigns and discovered eclipse modulations, transient dips and flares that may be related to accretion rate variations, two new orbital periods for known CVs and the confirmation of a new CV. These studies recognized that PDC data does not treat week-to-month trends accurately, and so only use the SAP light curves.

\subsection{X-ray Binaries}

X-ray binaries consist of a main sequence companion transferring mass by accretion onto a neutron star or black hole. The first such source discovered, Sco~X-1, fortunately fell within K2's Campaign 2 and was monitored with 1-minute cadence. \citet{Scaringi2015} and \citet{Hynes2016} found that the K2 light curve of Sco~X-1 is bimodal, indicating high and low states, and using simultaneous X-ray monitoring found that these states corresponded to periods of different X-ray behavior (namely, the object occupying the flaring and normal branches of the so-called Z-diagram). X-ray binaries are relatively rare objects, so opportunities for studying them with pre-determined fields of view were scarce until K2's Field 11, pointed at the Galactic Center. This field was observed in autumn of 2016, and several guest observers proposed for K2 monitoring of Galactic Center XRBs. We can therefore expect further XRB results in the very near future.

\subsection{Supernovae}

The final high-energy application to be discussed in this review is not accretion-based at all. For years, theoretical models of Type II supernovae predicted a flash before the main explosion caused by the core-collapse shock wave reaching and disrupting the stellar surface, called ``shock breakout." Additionally, information about the progenitor star can be gleaned from the rising light curve, since the profile of the rise in flux over time depends on the size and temperature of the expanding photosphere, which depend on the progenitor mass and radius. In Type Ia supernovae, which are caused by thermonuclear explosions on the surfaces of white dwarfs due to accretion of mass from a companion star, the early-time light curve depends on the density and velocity of the ejecta and the distribution of radioactive nickel. Of course, to observe early-time light curves of supernova one must be very lucky, since one would have to be looking in the right place at the right time to catch a supernova before it happens. With this in mind, several investigators (most members of the \textit{Kepler} ExtraGalactic Survey, or KEGS, team) selected a few hundred galaxies to monitor with \textit{Kepler}, in hopes that supernovae would go off. This was successful, and \citet{Olling2015} reported the pre-explosion light curves of three Type Ia supernovae, with early time behavior different from the few sparsely sampled early light curves already in existence. The authors emphasize that the new high cadence data enable new, much tighter constraints on models of explosion mechanisms. In a widely publicized result, \citet{Garnavich2016} detected a shock breakout flash in the pre-explosion \textit{Kepler} light curve of a Type II supernova, and for this object and one additional supernova were able to calculate the propagation time of the shock within the star and the progenitor radii to unprecedented precision. 

\section{TESS and Future Missions}

As of July 2018, the Transiting Exoplanet Survey Satellite or TESS is successfully taking data. This mission is the successor to \textit{Kepler} and CoRoT, but differs in many important ways that will affect its usefulness for high energy applications. 

\begin{itemize}
\item Sky coverage: The largest advantage that TESS will have over \textit{Kepler} for high energy applications will be its nearly all-sky coverage. While \textit{Kepler}'s exoplanet focus meant that it was originally pointed only near the galactic plane, preventing it from overlapping most extragalactic surveys, TESS is surveying the full sky with 30 minute cadence except for a narrow band around the ecliptic. The baselines vary from 27 days to 365 days, with objects at the ecliptic poles receiving the longest baselines (these are also in the JWST continuous viewing zone, for exoplanet characterization purposes). For AGN science, baselines shorter than $\sim$100 days are unlikely to yield robust characteristic timescales; however, shorter light curves could still characterize the high-$\nu$ PSD slope or variability amplitudes. The full sky coverage should be fantastic for early time supernova light curves.
\item Monitoring: Unlike with \textit{Kepler}, TESS will provide 30 minute cadence for every object in its very large sky coverage, regardless of whether or not a proposal was written. Proposals are only required for short cadence monitoring of guest observer targets. 
\item Bandpass: TESS has a single very wide optical bandpass, but the spectral coverage is redder than that of \textit{Kepler} by about 200 nm with good throughput even beyond 1$\mu$m. The redder color was chosen because TESS will primarily search for planets around M dwarf type stars, rather than solar type. It is possible that this will be advantageous, enabling AGN variability investigations out to higher redshifts since the UV/blue light from high-$z$~accretion disks will be varying in this band. However, the limiting magnitude should be considered carefully when selecting distant quasars for TESS monitoring.
\item Limiting magnitude: While \textit{Kepler} could nominally go down as far as 20th magnitude, we and most other studies found that objects fainter than $V\sim18.5$~were too noisy for meaningful analysis. TESS is predicted to achieve 1\% photometric precision at 16th magnitude, so TESS targets will need to be brighter than \textit{Kepler} targets. However, the much-increased sky coverage should make up for this in terms of total number of monitoring candidates.
\item Pixel size: Because \textit{Kepler} was focused on pointing stability and flux collection, it did not emphasize imaging. The \textit{Kepler} pixels are therefore large, about 4 arcsec. The TESS pixels will be even larger, at 21 arcsec. The main impact will be that crowded fields will become very challenging, since multiple varying targets are likely to be convolved. Methods do exist, however, for disentangling multiple varying sources (see, for example, the \textit{K2P}$^{2}$ pipeline discussed above and in Lund et al. 2015).

\end{itemize}

In 2019, ESA will launch the CHEOPS spacecraft to characterize exoplanets around bright stars. The limiting magnitudes are likely to be too bright for AGN, but may be appropriate for XRBs or CVs. Currently the mission plans to leave 20\% of its time available for guest observing. More interesting perhaps for high energy applications is PLATO, with a current estimated launch in 2026. Particularly attractive is the very fast cadence options, at 2.5 and 25 seconds. The baselines will be relatively long, from months to years, potentially appropriate for AGN applications. There is as yet little information on the limiting magnitude, but it appears to be approximately $V\sim15$, so some brighter quasars and AGN are likely to be viable. There will be guest observer program for this instrument.

\section{Conclusion}\label{sec5}

Many of the characteristics that enable missions like \textit{Kepler} and TESS to detect large numbers of exoplanetary transits can also enable powerful studies of accretion, explosions, and other time-dependent high energy processes. Only 4\% of publications coming from \textit{Kepler} and K2 concerned extragalactic targets, with similarly low percentages for accreting targets in the Galaxy. With a larger community aware of these unprecedented data, more rapid and collaborative progress can perhaps be made towards leveraging them.

\bibliography{xmm_proc_KLSmith_rev}
\end{document}